\documentclass[%
twocolumn,
superscriptaddress,
 amsmath,amssymb,
 aps,
]{revtex4-1}

\usepackage{mathtools}
\DeclarePairedDelimiter{\ceil}{\lceil}{\rceil}  
\usepackage{rotating}
\usepackage{graphicx}
\usepackage{subfigure} 
\usepackage{dcolumn}
\usepackage{bm}
\usepackage{float} 
\usepackage{mathrsfs} 
\usepackage{calligra}  
\DeclareMathAlphabet{\mathcalligra}{T1}{calligra}{m}{n} 
\DeclareFontShape{T1}{calligra}{m}{n}{<->s*[2.2]callig15}{} 
\newcommand{\scripty}[1]{\ensuremath{\mathcalligra{#1}}} 
\usepackage{tablefootnote} 
\usepackage{caption} 
\usepackage{siunitx}  
\usepackage{multirow}  
\usepackage{scrextend}
 
\usepackage{appendix}

\begin{document}

\preprint{APS/123-QED}
\title{New smoothly connecting open curves for modeling nucleosome-decorated DNA}

\author{Seyed A. Sabok-Sayr} 
\email{ \scriptsize{Corresponding author.}\\
\scriptsize{Email addresses: saboksayr@physics.rutgers.edu (Seyed A. Sabok-Sayr), wilma.olson@rutgers.edu (Wilma K. Olson).}}
\affiliation{Department of Physics and Astronomy, Rutgers University, Piscataway, NJ, U.S.A.}
\author{Wilma K. Olson}

\affiliation{Department of Chemistry and Chemical Biology,}

\affiliation{Center for Quantitative Biology,\\ Rutgers University, Piscataway, NJ, U.S.A.}

\date{September 27, 2019}

\begin{abstract}
We introduce an analytical method to generate the pathway of a closed protein-bound DNA minicircle.
This is a general method which can be used to connect any two open curves with well defined mathematical definitions 
as well as pairs of discrete systems found experimentally. We used this method to  describe the configurations of torsionally relaxed, 
$360$-base pair DNA rings with two evenly-spaced, ideal nucleosomes. We considered superhelical nucleosomal pathways with 
different levels of DNA wrapping 
and allowed for different inter-nucleosome orientations. We completed the DNA circles with the smooth connectors and 
studied the associated bending and electrostatic energies for different configurations in the absence and presence of salt. 
The predicted stable states bear close 
resemblance to reconstituted minicircles observed under low and high salt conditions. 

\end{abstract}

\maketitle

\section{Introduction}
\noindent

The two meters of DNA found in almost every human cell must be folded to fit in the nucleus of the cell ($\sim 6 \mu m$ in 
diameter \cite{MolBioCelText}). 
The first step in this compaction involves the coiling of $\sim150$ base pairs (bp) of negatively charged 
DNA around a core of eight positively charged histone proteins \textemdash two each of histones H2A, H2B, H3, and H4 \textemdash to form 
a nucleosome, the basic building block of chromatin \cite{McGheeAnnRev}. Nucleosome positioning determines the structure of chromatin and 
the expression 
which regulates the functional behavior of a cell.
Knowing how the superhelical stretches of DNA in nucleosomes are connected to one another is very important in understanding the folded 
structure of chromatin. Assuming the path of DNA in each nucleosome as a three-dimensional $(3D)$  open curve, the problem becomes how to connect 
the end of one curve smoothly to the end of the next.
In this paper we introduce an analytical method to connect any two nearby well 
defined  $3D$ open curves smoothly. This is a general method which can be applied to connect curves of interest with well 
defined mathematical 
definitions as well as discrete systems found experimentally, such as the sets of atomic coordinates determined by 
X-ray crystallography or cryo-electron microscopy. Without loss of generality we can assume that the smooth connector has a constant slope 
along an arbitrary direction, 
taken here to be the cylindrical axis ($z$-axis) of one of the nucleosomes. We formulate 
a $3D$ smooth pathway by decomposing 
the trajectory of the connector along the $z$-axis and onto a two-dimensional $(2D)$ $xy$-plane. We find that the simplest possible 
smooth connector that can satisfy the boundary conditions of the end points of two open 
curves has a quartic polynomial trajectory on the $2D$ plane. Here we consider smooth pathways of B DNA in which the centers of base pairs 
align with the DNA symmetry axis and  fall along the trajectory of the curve. The combined set of nucleosomal DNA pathways and 
connector curves thus describe a relaxed, twist-free DNA system.

We applied this approach to small nucleosome-decorated DNA minicircles similar to chromatin fragments previously characterized by 
electron microscopy under low and hight salt conditions \cite{Goulet}. We compared the predicted energies of the nucleosome-decorated 
DNA to the shapes determined experimentally. We considered ideal, cylindrically shaped nucleosomes with different levels of DNA wrapping 
and allowed for different orientations of the nucleosomes with respect to one another.
We ranked the configurations in terms of their elastic and electrostatic energies. We considered the electrostatic contribution in the 
presence and absence of salt, and given that the generated DNA configurations are torsionally relaxed and the chain is assumed to be 
inextensible, we only determined the bending contribution to the elastic energy.
We find stable configurations of DNA that resemble the experimental observations. Our findings suggest that the uptake of salt may 
increase the wrapping of DNA on the nucleosome.

\section{Theory}
Suppose we have two nearby open curves 
$\mathscr{C}_1$ and $\mathscr{C}_2$, and we want to  connect one end of  $\mathscr{C}_1$ located at $\scripty{r}_1$ with slope 
$\scripty{m}_1$ smoothly to an end of $\mathscr{C}_2$ located at $\scripty{r}_2$ with slope $\scripty{m}_2$. For a connector $\ell$ to be smooth
it should start from curve  $\mathscr{C}_1$ at $\scripty{r}_1$ with $\scripty{m}_1$ and end at curve $\mathscr{C}_2$ at 
$\scripty{r}_2$ with $\scripty{m}_2$. Without loss of generality we can assume that $\ell$ has a constant slope along an arbitrary direction, 
taken here to be the $z$-axis. This assumption reduces the problem from finding $\ell$ in $3D$ to finding the projection of $\ell$ onto a $2D$ 
plane, i.e., 
$\ell_{xy}$ on the $xy$-plane, and also makes it possible to connect the arc length of 
 $\ell_{xy}$ to the arc length of $\ell$ in $3D$ by using the Pythagorean theorem.
 
 We are going to develop an expression for the total
 arc length, or contour length, of the connector and for the coordinates of points at specified fractional contour lengths.
 Given that the length of a curve is invariant to rotation and translation, the result will be 
valid in any
 coordinate system. So the problem is to find an equation of a curve $y(x)$ in $2D$ which starts from an initial position, 
$ r_i=(x_i,y_i)$ with initial slope $r'_i = r'(x_i, y_i) = y'(x_i)$, and ends 
at a final position, $r_f=(x_f,y_f)$ with final slope $r'_f =r'(x_f,y_f) = y'(x_f)$. 
There is also a length constraint, which is the minimum value that must be satisfied by the equation of the connector. Depending
on the problem, the contour can be either continuous or discrete. For instance, if the connector is used to describe a DNA pathway, we need to incorporate
its discrete length $S$ in our treatment. The value of $S$ is determined by the number of base pairs $N$ and the distance $s=3.4\;\si{\AA}$  
between successive base pairs, i.e., $S=Ns$. Therefore the function $y(x)$ should satisfy five conditions$:$ two boundary 
conditions for the initial and final positions, two for the initial and final slopes, and a constraint on the length. 
The equation should also be transferable from one 
coordinate system to another. Therefore the function $y(x)$ must be
well defined and differentiable within $(-\infty, +\infty)$. Given these constraints, the simplest representation of a smooth connector
is a polynomial of order four with five coefficients: 

\begin{equation}
\resizebox{.9\hsize}{!}{
$y(x)= a_4 x^4 + a_3 x^3 + a_2 x^2 + a_1 x + a_0\;\;\;\;\;\;\;\;\; x_i \leq x \leq x_f. $
}
\end{equation}

\subsection{Connector contour length}
The constraints on the ends and length of the connector uniquely define the five coefficients $a_4, a_3, a_2, a_1,$ 
and $a_0$ in Eq.$(1)$, 
which in turn uniquely define the smooth connecting function $y(x)$. The four boundary conditions 
\textemdash $\,$ i.e., $y_i = y(x_i),$ $y_f = y(x_f),$ 
$y'_i = y'(x_i),$ and 
$y'_f = y'(x_f) \,$ \textemdash $\,$ lead to a set of four equations with five unknowns:  
\begin{equation}
 \begin{cases}
  & a_4 x_i^4 + a_3 x_i^3 + a_2 x_i^2 + a_1 x_i + a_0 = y_i \\
\vspace{1mm}  
  & a_4 x_f^4 + a_3 x_f^3 + a_2 x_f^2 + a_1 x_f + a_0 = y_f \\
\vspace{1mm}   
  &4 a_4 x_i^3 +3 a_3 x_i^2 + 2 a_2 x_i + a_1  = y_i'\\
\vspace{1mm}   
  & 4 a_4 x_f^3 + 3 a_3 x_f^2 + 2 a_2 x_f + a_1  = y_f' \;\; .
 \end{cases}
\end{equation}

By treating $a_4, a_3, a_2, a_1,$ and $a_0$ as five unknowns and the different values of 
$x_i, x_f, y_i, y_f, y'_i,$ and $y'_f$ as known coefficients, we 
can use Gauss-Jordan elimination to find four of the unknown coefficients in terms of the fifth one, e.g., $a_3, a_2, a_1,$ 
and $a_0$ as linear functions of $a_4$,
\begin{equation}
\resizebox{.9\hsize}{!}{$
\begin{bmatrix}
 x_i^4   &   x_i^3   & x_i^2 & x_i  & 1  &   y_i\\
 x_f^4   &   x_f^3   & x_f^2 & x_f  & 1  &  y_f \\
 4x_i^3  &   3x_i^2  & 2x_i  &   1  & 0  &   y'_i\\
 4x_f^3  &   3x_f^2  & 2x_f  &   1  & 0  &   y'_f 
\end{bmatrix}
\Longrightarrow 
\left\{
\begin{aligned}
 a_3 &= c_{3a} a_4 + c_{30}  \\
 a_2 &= c_{2a} a_4 + c_{20}  \\
 a_1 &= c_{1a} a_4 + c_{10}  \\
 a_0 &= c_{0a} a_4 + c_{00}  \;\; ,
\end{aligned}\right.
$}
\end{equation}
where $c_{3a}, c_{30}, c_{2a}, c_{20}, c_{1a}, c_{10},  c_{0a}, $ and $c_{00}$ are expressed as follows: 
\begin{equation*}
c_{3a} = -2(x_f + x_i)\;\; , 
\end{equation*}
\begin{equation*}
c_{30} = - \dfrac{2(y_f - y_i)+(x_i-x_f)(y'_f+y'_i)}{(x_f-x_i)^3}\;\; ,
\end{equation*}
\begin{equation*}
c_{2a} = -(x_i^2+4x_i x_f + x_f^2) \;\; ,
\end{equation*}
\begin{equation*}
 \begin{split}
 c_{20} = & - \dfrac{1}{(x_f - x_i)^3} (-3 x_f y_f - 3 x_i y_f + 3 x_f y_i + 3 x_i y_i \\ 
  & + x_f^2 y'_f + x_f x_i y'_f - 2 x_i^2 y'_f + 2 x_f^2 y'_i - 
   x_f x_i y'_i - x_i^2 y'_i) \;\; ,
 \end{split}
\end{equation*}
\begin{equation*}
c_{1a} = -2 (x_i^2 x_f + x_i x_f^2) \;\; ,
\end{equation*}
\begin{equation*}
\begin{split}
c_{10} =& -\dfrac{1}{(x_f - x_i)^3} (6 x_f x_i y_f - 6 x_f x_i y_i - 2 x_f^2 x_i y'_f \\  
 & + x_f x_i^2 y'_f + x_i^3 y'_f - x_f^3 y'_i - x_f^2 x_i y'_i + 2 x_f x_i^2 y'_i) \;\; , 
 \end{split}
\end{equation*} 
\begin{equation}
c_{0a} = x_i^2 x_f^2 \;\; ,
\end{equation}
\begin{equation*}
\begin{split}
c_{00} =&  -\dfrac{1}{(x_i-x_f)^3} (3 x_f x_i^2 y_f - x_i^3 y_f + 
x_f^3 y_i - 3 x_f^2 x_i y_i \\
   &- x_f^2 x_i^2 y'_f + x_f x_i^3 y'_f - x_f^3 x_i y'_i + 
   x_f^2 x_i^2 y'_i) \;\; .
\end{split}
\end{equation*}

Upon substitution of the coefficients $a_3, a_2, a_1,$ and $a_0$ in the expression for the connector in $2D$, $y$ becomes a function of both $x$ and 
the parameter $a_4$,
\begin{equation}
\resizebox{.9\hsize}{!}{$
\begin{split}
y = y(x,a_4) = &\,a_4 x^4 + ( c_{3a} a_4 + c_{30})x^3 + ( c_{2a} a_4 + c_{20} )x^2 \\
               & + ( a_{1a} a_4 + a_{10}) x  + a_{0a} a_4 +  a_{00}\;\; .
\end{split}
$}
\end{equation}

The length of the curve that satisfies the four boundary conditions in $2D$ is also a function of the parameter $A$:\\ 
\begin{equation}
\resizebox{1.1\hsize}{!}{$  
\begin{split}
 S^{2D}(a_4) &= \int_{x_i}^{x_f}\sqrt{1+\Big[y'(x,a_4)\Big]^2} dx \\
         &= \int_{x_i}^{x_f} \sqrt{1 + \Big[4a_4x^3 + 3(c_{3a}a_4 + c_{30})x^2 + 2( c_{2a}a_4 + c_{20} )x + (c_{1a}a_4 + c_{10}) \Big]^2 }\;\;dx \;\; .
 \end{split}
 $}
\end{equation}

The minimum length of the trajectory in the $xy$-plane, is found by first determining the value of $A$ for which the derivative of 
$S^{2D}$ vs. $a_4$ vanishes, 
\begin{equation}
\resizebox{.5\hsize}{!}{$
\dfrac{\partial S^{2D}}{\partial a_4} = \displaystyle \int_{x_i}^{x_f} \dfrac{y' \bigg(\dfrac{\partial y'}{\partial a_4}\bigg)}{\sqrt{1+y'^2}}dx =0\;\; . 
$}
\end{equation}

In general, Eq. $(7)$ may have two numerical solutions, positive and negative: $A_{opt}^+$ and $A_{opt}^-$, corresponding to two 
possible smooth connectors. Depending on the boundary conditions the two solutions may converge to one. A system with
$N$ connections will in general have $2^N$ possible combinations of connections.

The value(s) of $a_4$ found upon solution of Eq.$(7)$, $a_4^{opt},$ can then be used to obtain the minimum theoretical value(s) of 
the length along the connector(s) in $2D$, $S^{2D}$, as:
\begin{equation}
\resizebox{.9\hsize}{!}{$
\displaystyle S_{min}^{2D-theory} = S^{2D}(a_4^{opt}) = \int_{x_i}^{x_f}\sqrt{1+\Big[y'(x,a_4^{opt})\Big]^2} dx\;\; .
$}
\end{equation}

Since we assumed that the slope of the connector in the $z$-direction is constant, the value of $S_{min}^{2D-theory}$ and the minimum theoretical 
length of the connector in $3D$, $S_{min}^{3D-theory}$, are related by the Pythagorean theorem:
\begin{equation}
S_{min}^{3D-theory} = \sqrt{(S_{min}^{2D-theory})^2 + (z_f - z_i)^2} \;\; .
\end{equation}

\subsection{Continuous connector}

The value of $S_{min}^{3D-theory}$ is the minimum length required for a continuous connector to satisfy
the initial and final boundary conditions. If the problem requires a connector with a length greater than the minimum length, i.e.,
$S^{3D-continuous} > S_{min}^{3D-theory}$, one should first find the projection of the larger curve in $2D$, $S^{2D-continuous}$, by 
the Pythagorean theorem: 
\begin{equation}
\resizebox{.85\hsize}{!}{$
S^{2D-continuous} = \sqrt{(S^{3D-continuous})^2 - (z_f - z_i)^2} \;\; .
$}
\end{equation}

The parameter $a_4^\text{*}$ associated with this length can be obtained by solving Eq.$(6)$:
\begin{equation}
\begin{split}
 S^{2D}(a_4^\text{*}) &= \int_{x_i}^{x_f}\sqrt{1+\Big[y'(x,a_4^\text{*})\Big]^2} dx \;\; .
\end{split}
\end{equation}

After finding the value of $a_4^\text{*}$ associated with the projected curve of greater length, we can use Eq.$(3)$ 
to find the values of $a_3, a_2, a_1,$ and $a_0$. These parameters fully specify the equation of the longer connector in $2D$, i.e., Eq.$(1)$. The same
procedure can be used to determine the equation of the connector of minimum length.

As noted above, the length along the connector $\ell$ in $3D$ is related to that of its projection $\ell_{xy}$ on the $xy$-plane.  
Thus for every $x\in (x_i, x_f)$ there is a corresponding contour length along $\ell$ in $3D$:
\begin{equation}
 \displaystyle \ell(x) = \sec( \gamma ) \int_{x_i}^x\sqrt{1+\Big[y'(x)\Big]^2} dx\;\; ,
\end{equation}
where $\gamma$ is a constant angle determined by the slope of $\ell$ in the $z$-direction:
\begin{equation}
 \gamma = \cos^{-1}\left(\dfrac{S^{2D-continuous}}{S^{3D-continuous}}\right).
\end{equation}

The equation of the connector $\ell$ in $3D$ can then be expressed as:
\begin{equation}
\left\{
\begin{split}
  x &= x  \;\;\;\;\;\;\;\;\;\;\;\; x_i \leq x \leq x_f  \\
  y &= a_4x^4 + a_3x^3 + a_2x^2 + a_1x + a_0\;\;\; \\
  z &= z_i + \dfrac{\ell(x)}{S^{3D-continuous}} (z_f -z_i) \;\; . 
\end{split} \right.
\end{equation}

\subsection{Discrete connector}

The connector for a discrete system such as DNA is not necessarily equal to $S_{min}^{3D-theory}$. The value of $S_{min}^{3D-theory}$ can 
be any real number, not necessarily the length of the discrete system. The length of a discrete connector is in general described as: 
\begin{equation}
 S =   n \delta \;\;,
\end{equation}
 where $n$ is the number of repeating units which comprise the connector and $\delta$ is the length  of each repeating unit. 
 In order to treat the 
 discrete system, we need to find the minimum number $n_{min}$ of repeating units in the system that is compatible with the minimum 
 theoretical contour length. In order to find $n_{min}$, we divide 
  $S_{min}^{3D-theory}$ by  the value of $\delta$. Since the quotient  is not necessarily an integer, we round the quotient up
  to the nearest integral value. That is to say we find the ceiling of the quotient i.e., 
\begin{equation}
 n_{min} = \ceil*{\dfrac{S_{min}^{3D-theory}}{\delta}}\;\; . 
\end{equation}

We then use $n_{min}$ to find the actual minimum discrete length which satisfies the boundary conditions: 
\begin{equation}
 S_{min}^{3D-discrete} = n_{min}  \delta \;\; .
\end{equation}

We then use the Pythagorean theorem to find $S_{min}^{2D-discrete}$ : 
\begin{equation}
 S_{min}^{2D-discrete} = \sqrt{(n_{min}  \delta)^2 - (z_f - z_i)^2} \;\; .
\end{equation}

The value of the parameter $a_4$ associated with the minimum discrete length can be obtained by solving Eq.$(6)$:
\begin{equation}
\begin{split}
 S_{min}^{2D-discrete}(a_4) &= \int_{x_i}^{x_f}\sqrt{1+\Big[y'(x,a_4)\Big]^2} dx\;\;.
\end{split}
\end{equation}

If the problem requires a discrete connector with a length greater than the minimum length, i.e.,
$n > n_{min}$, one should first find $S^{2D-discrete}$ by the Pythagorean theorem:
\begin{equation}
 S^{2D-discrete} = \sqrt{(n \delta)^2 - (z_f - z_i)^2} \;\; ,
\end{equation}
then the parameter $a_4^\text{*}$ can be found by solving Eq.$(6)$:
\begin{equation}
\begin{split}
 S^{2D-discrete}(a_4^\text{*}) &= \int_{x_i}^{x_f}\sqrt{1+\Big[y'(x,a_4^\text{*})\Big]^2} dx\;\;.
\end{split}
\end{equation}

After finding the value of $a_4^\text{*}$ associated with the projected curve of greater length, we can use Eq.$(3)$ 
to find the values of $a_3, a_2, a_1,$ and $a_0$. These parameters fully specify the equation of the longer connector in $2D$, i.e., Eq.$(1)$. The same
procedure can be used to determine the equation of the connector of minimum length.

As noted above, the length along the connector $\ell$ in $3D$ is related to that of its projection $\ell_{xy}$ on the $xy$-plane. 
 Thus, every discrete contour length along $\ell$ in $3D$, $j \delta$ for a $j\in [0, n]$, there is a corresponding 
 $x_j \in [x_0, x_n]$ which can be found by solving the following integral for $x_j$:
\begin{equation}
 \displaystyle \ell(x_j)= j \delta = \sec( \gamma ) \int_{x_0}^{x_j}\sqrt{1+\Big[y'(x)\Big]^2} dx\;\; ,
\end{equation}
where $\gamma$ is a constant angle determined by the slope of $\ell$ in the $z$-direction:
\begin{equation}
 \gamma = \cos^{-1}\left(\dfrac{S^{2D-discrete}}{S^{3D-discrete}}\right).
\end{equation}

The equation of the connector $\ell$ in $3D$  can then be expressed as:
\begin{equation}
\left\{
\begin{split}
  x_j &= x_j  \;\;\;\;\;\;\;\;\;\;\;\; 0 \leq j \leq n  \\
  y_j &= a_4x_j^4 + a_3x_j^3 + a_2x_j^2 + a_1x_j + a_0\;\;\; \\
  z_j &= z_i + \dfrac{j}{n} (z_f -z_i) \;\; .
\end{split} \right.
\end{equation}

\subsection{DNA Model}
\noindent
\textbf{Nucleosome-bound DNAs.}  We model DNA rings containing two cylindrically shaped ideal nucleosomes. The nucleosome-bound portions of
DNA are represented by left-handed circular superhelices and the intervening, protein-free linker DNA by smooth connectors (Eq.$(24)$).
The center of one nucleosome is taken to lie 
at the origin with its cylindrical axis along the global $z$-axis. The nucleosome is oriented such that its dyad axis runs parallel to the 
global $x$-axis. The coordinates of the DNA on the reference nucleosome are thus:
\begin{equation}
\left\{
\begin{split}
  x & = r \cos \theta  \\
  y & = r \sin \theta\\
  z & = - p \theta,
\end{split} \right.
\end{equation}
\noindent
where $r$ is the radius of the superhelix, $p$ is the pitch, and $\theta$ is the cylindrical rotation of DNA about the superhelical axis. 
The value of $\theta$
ranges from an initial angle of $\theta_i$ to a final angle of $\theta_f$, which respectively correspond to initial and final coordinates 
of the superhelix, i.e., the locations of the centers of the first and last nucleosome-bound base pairs. 

The second nucleosome is separated from the centers of the first one by a distance $d$ along the global $x$-axis and is rotated 
by an angle $\alpha$ about the same axis. The coordinates of the second nucleosome-bound DNA are:
\begin{equation}
\left\{
\begin{split}
  x & = r \cos \theta + d \\
  y & = r \sin \theta \cos \alpha - p \theta \sin \alpha \\
  z & = - r \sin \theta \sin \alpha - p \theta \cos\alpha ,
\end{split} \right.
\end{equation}
\noindent
where $\theta$ is the cylindrical rotation of the second nucleosome about its superhelical axis. The value of $\theta$ ranges from initial
 and final angles corresponding to initial and final coordinates of the second superhelix. The choice of initial and final values determines the 
 orientation of the second nucleosome with respect to the first. 
 
 The equation of a smooth connector, i.e., Eq.$(24)$, provides a 
 representation of a flexible, protein-free pathway of the linker DNA  
 that connects the terminus of one nucleosome to the start of another nucleosome. Two such connectors are required to form the desired 
 nucleosome-DNA assembly. Here we assume connectors of the same length leading to evenly spaced nucleosomes. Under certain conditions
 the two connectors may self-intersect. Since it is not physically possible that the two connectors pass 
 into one another,
 we set the coordinates and the tangents of both connectors at the crossing point as the boundary conditions of new midway points along 
 each connector.  
 We then use the initial boundary conditions and the midway conditions to split each initial connector into two parts. 
 Knowing the equations of the nucleosomal and protein-free DNA we can find the coordinates of the points representing the
centers of the base pairs along the DNA assuming that the double helix adopts the B form. \\

\noindent
\textbf{Bending Energy.} We consider the deformations of DNA resulting from its interaction with proteins, i.e. the wrapping
of the double helical structure along a superhelical pathway, and the molecular distortion required to connect successive
nucleosomes.
We measure the deformation of DNA compared to a naturally straight, inextensible, linearly elastic, isotropic rod
with circular cross section. The energy associated with deformation of such a rod is expressed as a sum of bending and 
 twisting contributions. The minicircle is assumed here to have at least one single-stranded scission 
 and is thus torsionally relaxed. The energy associated with deformation is then simply the bending energy. 
 
 From the equations of the smooth connector and the circular superhelix, we can determine the components of 
 the tangent to the curve at 
any point. Using the components of the tangents we can calculate the angle of bending $\bigtriangleup \eta_i$ at each base-pair step, i.e.,
$\bigtriangleup \eta_i = \cos^{-1}\left(\frac{t^i \cdot t^{i+1}}{|t^i| |t^{i+1}|} \right) $ where $t^i$ is the tangent vector, $|t^i|$ 
is its magnitude, $i=1-N$, and $N$ is the total number of base pairs. 
The total bending energy $\Psi$ associated with the configuration of the DNA is then given by:
\begin{equation}
 \Psi = \dfrac{1}{2}k_{\text{B}}T \sum_i \left(\dfrac{\bigtriangleup\eta_i^2}{\langle\bigtriangleup\eta^2\rangle} \right)\; ,
\end{equation}
\noindent
where  the bending stiffness  
of individual base-pair steps is consistent with the known $\sim 50\si{nm}$ persistence length of DNA \cite{e}. The energy is raised 
by $\frac{1}{2}k_{\text{B}}T$ when the direction 
of a step deviates from its equilibrium rest state by its root-mean-square fluctuation $\langle\bigtriangleup\eta^2\rangle^{1/2}$, 
here taken to be $4.82^\circ$.\\

\noindent
\textbf{Electrostatic Energy.} We also estimate the electrostatic energy by considering the interactions of 
the negatively charged phosphate groups on the DNA backbone. The two charges associated with each base pair are merged into a single charge of 
twice the magnitude and placed, for simplicity, at the center of the base pair. The electrostatic energy of DNA is taken to be the sum of all pairwise 
interactions between the charges on different base pairs:
\begin{equation}
\Phi=\sum_{j=2}^N \sum_{i=1}^{j-1}\dfrac{\delta q_i \delta q_j e^{-\kappa r^{ij}}}{4\pi \epsilon_r \epsilon_0 r^{ij}}\; , 
\end{equation}
where $\delta q_i$ and $\delta q_j$ are the respective charges associated with the $ith$ and $jth$ base pairs, 
$r^{ij}$ is the distance between the centers of those base pairs, $\epsilon_r$ is the relative permitivity of water
at $300\,K$ ($\sim 80$), and 
$\kappa = 0.329\sqrt{C_m}$ is the Debye screening parameter for monovalent salt, such as NaCl,  of molar 
concentration $C_m$. For protein-free DNA, we assume $76\%$ charge neutralization associated with the screening of DNA charges by counterions \cite{d}, 
corresponding to $2\times 0.24 e^- = 7.70\times 10^{-20}$C per base pair. For protein-bound DNA, we assume $86\%$ charge neutralization 
associated with the combination of counterions and the net charge of the nucleosome, corresponding to  $2\times 0.14 e^- = 4.49\times 10^{-20}$C 
per base pair as found in molecular dynamic (MD) simulation of nucleosomes \cite{h}. \\

\noindent
\textbf{Writhing Number.} The writhing number $Wr(C)$ is a topological property of a curve $C$ which is invariant under translation 
and rotation and measures the chiral distortion of the curve. A $3$-D curve has different projections
when viewed from different angles and its $Wr$ is the average of the number of positive and negative self-crossings over 
all projections \cite{f}.

\noindent
The writhing number of the nucleosome-DNA assembly is computed here using the formulation of Swigon et al \cite{g} for a closed discrete 
curve: 
\begin{equation}
 Wr = \dfrac{1}{2\pi}\sum_{(i,j),j>i}w^{ij}\;\; ,
\end{equation}
where: 
\begin{equation}
 w^{ij} = \xi^{i,j} + \xi^{i+1,j+1} - \xi^{i,j+1} - \xi^{i+1,j} \;\;,
\end{equation}
\noindent
and $\xi^{i,j}$ is the solution of the following pair of equations:
\begin{equation}
\left\{
\begin{split}
 \cos \xi^{i,j} & =  \dfrac{r^{ij}\times t^j}{|r^{ij}\times t^j|}\cdot \dfrac{t^i \times r^{ij}}{|t^i \times r^{ij}|}  \\
 \sin \xi^{i,j} & = -\dfrac{r^{ij}}{|r^{ij}|} \cdot\left( \dfrac{r^{ij}\times t^j}{|r^{ij}\times t^j|}\times \dfrac{t^i \times r^{ij}}{|t^i \times r^{ij}|}\right)\;\; . 
\end{split} \right.
\end{equation}

\subsection{Results}

We generated analytical and computational representations of a torsionally relaxed, $360$\nobreakdash-bp DNA ring with two evenly-spaced 
ideal nucleosomes. 
We considered nucleosomes with different levels of DNA wrapping, $1.5$ and $1.75$ superhelical turns, and allowed for different 
inter-nucleosome orientation angles $\alpha$ 
over the range $0^\circ$ to $90^\circ$. 
We  completed the DNA circle with two (or four) linkers defined
by the smooth connectors in Eq.$(24)$. We then calculated the bending energy, the electrostatic energy, and the writhing number of 
the DNA for each configuration. In our initial calculations the molar concentration $C_m$ of monovalent counterions was taken to be $0.1M$.

Table I reports the values of the coefficients $(a_{\mu k}, \mu=0-4, k=1-2)$ describing the two smooth connectors for minicircles 
containing two nucleosomes, each with
 $1.5$ turns of DNA, and oriented at different values of $\alpha$. As $\alpha$ changes the boundary conditions of the two protein-bound 
portions of the 
DNA minicircle change. Therefore for
every value of the angle $\alpha$ we get unique values for $(a_{\mu k}, \mu=0-4, k=1-2)$. The reported values of $d$ in the table 
are the shortest distances between nucleosome centers that satisfy the boundary conditions.
\begin{table}[H]
\hfil                 
\begin{turn}{90}
   \begin{minipage}{1.55\linewidth}
   \begin{center}
      \caption{\small{Minimum separation distances and coefficients$^\dagger$ of smooth connectors between two evenly spaced nucleosomes, 
      each with $1.3$ turns of DNA, on a torsionally relaxed, $360$-bp minicircle as a 
      function of the orientation angle $\alpha$.}}
      \scalebox{0.68}{%
      \label{tab:table1}
      \begin{tabular}{|c|c|c|c|c|c|c|c|c|c|c|c|}
       \hline
       \textbf{$\alpha\;(^\circ$)} & \textbf{$d$($\si{\AA}$)} & \textbf{$a_{41}(\si{\AA^{-3}})$} & \textbf{$a_{31}(\si{\AA^{-2}})$} & \textbf{$a_{21}(\si{\AA^{-1}})$} & \textbf{$a_{11}$} &
       \textbf{$a_{01}(\si{\AA})$} & \textbf{$a_{42}(\si{\AA^{-3}})$} & \textbf{$a_{32}(\si{\AA^{-2}})$} & \textbf{$a_{22}(\si{\AA^{-1}})$} & \textbf{$a_{12}$} &
       \textbf{$a_{02}(\si{\AA})$}   \\   
       \hline
        $0$ &  $200.94$   & $1.61\times10^{-7}$ & $6.47 \times 10^{-5}$  & $6.47 \times 10^{-3}$   & $-4.42 \times 10^{-3}$  & $43.29$  & 
        $-1.61\times 10^{-7}$ & $-6.47 \times 10^{-5}$  &  $-6.47 \times 10^{-3}$   &  $4.42 \times 10^{-3}$ & $-43.29$  \\
        $15$ &  $202.64$   & $1.54\times 10^{-7}$ & $6.15 \times 10^{-5}$  & $6.05 \times 10^{-3}$   & $-5.85 \times 10^{-3}$  & $46.24$  & 
        $-1.54\times 10^{-7}$ & $-6.15 \times 10^{-5}$  &  $-6.05 \times 10^{-3}$   &  $5.85 \times 10^{-3}$ & $-46.24$   \\
        $30$ &  $203.66$   & $1.50\times 10^{-7}$ & $6.01 \times 10^{-5}$  & $5.93 \times 10^{-3}$   & $-7.37 \times 10^{-3}$  & $46.04$  & 
        $-1.50\times 10^{-7}$ & $-6.01 \times 10^{-5}$  &  $-5.93 \times 10^{-3}$   &  $7.37 \times 10^{-3}$ & $-46.04$   \\
        $45$ &  $203.66$   &  $1.56\times 10^{-7}$ & $6.33 \times 10^{-5}$  &  $6.41 \times 10^{-3}$  &  $-9.08 \times 10^{-3}$ & $42.70$  &
        $-1.56\times 10^{-7}$ & $-6.33 \times 10^{-5}$  & $-6.41 \times 10^{-3}$  & $9.08 \times 10^{-3}$  & $-42.70$   \\
        $60$ &  $203.66$   &  $1.44\times 10^{-7}$ & $6.02 \times 10^{-5}$  &  $6.38 \times 10^{-3}$  &  $-1.01 \times 10^{-2}$ & $36.46$  &
        $-1.44\times 10^{-7}$ & $-6.02 \times 10^{-5}$  & $-6.38 \times 10^{-3}$  & $1.01 \times 10^{-2}$  & $-36.46$   \\
        $75$ &  $202.64$   &   $1.42\times 10^{-7}$ & $6.10 \times 10^{-5}$  &  $6.85 \times 10^{-3}$  &  $-1.11 \times 10^{-2}$ & $27.72$  &
        $-1.42\times 10^{-7}$ & $-6.10 \times 10^{-5}$  & $-6.85 \times 10^{-3}$  & $1.11 \times 10^{-2}$  & $-27.72$  \\
        $90$ &  $200.94$   &    $1.40\times 10^{-7}$ & $6.23 \times 10^{-5}$  &  $7.46 \times 10^{-3}$  &  $-1.17 \times 10^{-2}$ & $17.10$  &
        $-1.40\times 10^{-7}$ & $-6.23 \times 10^{-5}$  & $-7.46 \times 10^{-3}$  & $1.17 \times 10^{-2}$  & $-17.10$  \\
       \hline
      \end{tabular}}  
      \captionsetup{justification=justified, singlelinecheck=off}
      \caption*{\footnotesize{$^\dagger$The first subscript of each coefficient is its order (Eq.$(1)$) and the second one refers to the order of connectors.
      In the second subscript, values $1$ and $2$ refer respectively to the connectors joining the terminus of the second nucleosome 
      to the start of the first nucleosome and the terminus of the first nucleosome to the start of the second. }}

   \end{center}
   \end{minipage}  
\end{turn}
\end{table}
Figure 1 presents the bending energy $\Psi$ and electrostatic energy $\Phi$ of a series of nucleosome-decorated DNA minicircles as a function of the angle 
$\alpha$. 
The minimum bending energy occurs when $\alpha = 60^\circ$ and the minimum electrostatic energy when
$\alpha = 75^\circ$. Variation in the bending energy is greater than that in the electrostatic energy for the assumed choice of parameters 
($\epsilon_r=80$, $\kappa=0.104$, $74\%$ charge neutralization on DNA linkers, and $86\%$ neutralization on nucleosomal DNA) and thus determines the 
location of the minimum value of the total energy. 
It is possible that for different values of $\epsilon_r$, $\kappa$, and charge neutralization  
the electrostatic energy can overwhelm the bending energy and shift the angle of minimum energy to a higher value, e.g., less neutralization, 
smaller $\epsilon_r$, and/or smaller $\kappa$. In this figure we 
present the two energies in different scales to show that they attain their minima at different values of $\alpha$. On the top of every 
column is an image showing a top-down view of the upper (first, i.e., Eq. ($25$)) nucleosome, with its cylindrical axis coincident with the 
global $z$-axis of the system. As evident from the images, in the state of minimum bending energy the two connectors, on average, follow a 
straighter pathway and thus a smaller 
bending energy is stored in the configuration. Since the superhelical structure of the nucleosomal DNA does not change when the angle $\alpha$ 
changes, the bending energy associated with both nucleosomal DNAs is constant and therefore makes no contribution to 
the change in total bending energy. The pathway of both connectors, however, changes with $\alpha$. As expected, 
the average bending energy per base pair does not change for the nucleosomal DNA with change in $\alpha$ 
but the values of the bending energy for the connectors do change (see Table S-I). The energies of the two connectors are equivalent due to 
the symmetry of the system.
 \begin{figure}[H]
   \centering
   \caption{\small{Molecular images illustrating the changes in overall folding and graphs of the associated bending $\Psi$ and electrostatic 
   $\Phi$ energies  of a torsionally relaxed, $360$-bp DNA minicircle with two evenly spaced nucleosomes, each wrapping $1.5$ turns of DNA, 
   as a function of the angle $\alpha$ between nucleosome axes. The two dark gray cylinders in each molecular image represent the histones proteins 
   wrapped by DNA. See Table S-II for numerical values.}}
   \includegraphics[width=0.47\textwidth]{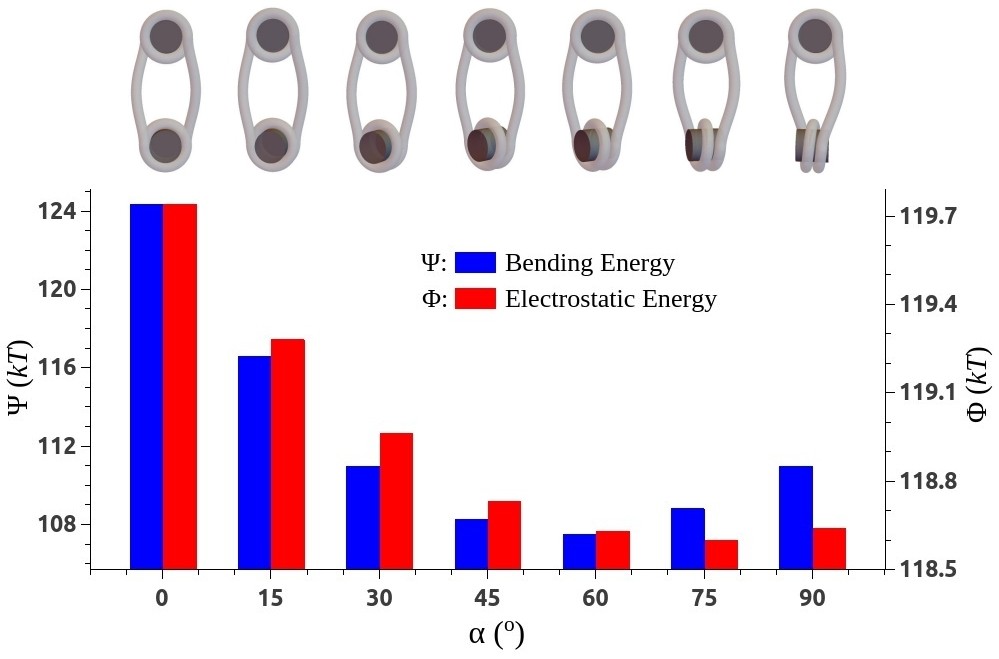}\\
\end{figure}

Figure 2 depicts the writhing number $Wr$ of the same nucleosome-decorated DNA along with two sets of molecular images. The two sets correspond
to different views of the same molecular images shown in Fig. $1$. The lower set in Fig. $2$ are side views obtained by rotating the 
images in Fig. 1 by $-90^\circ$ about the global $x$-axis, i.e., the long axis of the system parallel to the downward direction of the page. 
The resulting view is that along the global $+y$-axis with the first nucleosome remaining on the top. The upper set of images in Fig. $2$ 
are front views obtained by rotating the images in Fig. 1 by $-90^\circ$ about the global $y$-axis. The viewing direction then lies along the 
global $-x$-axis with the first nucleosome on top and in the back. 

As evident from the plotted values, the magnitude of the writhing number increases monotonically with increase in 
$\alpha$. The increase in magnitude corresponds to a greater number of self crossings in the chain. The added 
crossings are not evident in the (lower) side views in the figure nor in those in Fig. 1. The front views, however, clearly show the increase in 
self crossings when $\alpha$ increases. 

 \begin{figure}[H] 
   \centering
   \caption{\small{The writhing number of the nucleosome-DNA assemblies presented in Fig. 1. See Table S-II for numerical values.}}
   \includegraphics[width=0.45\textwidth]{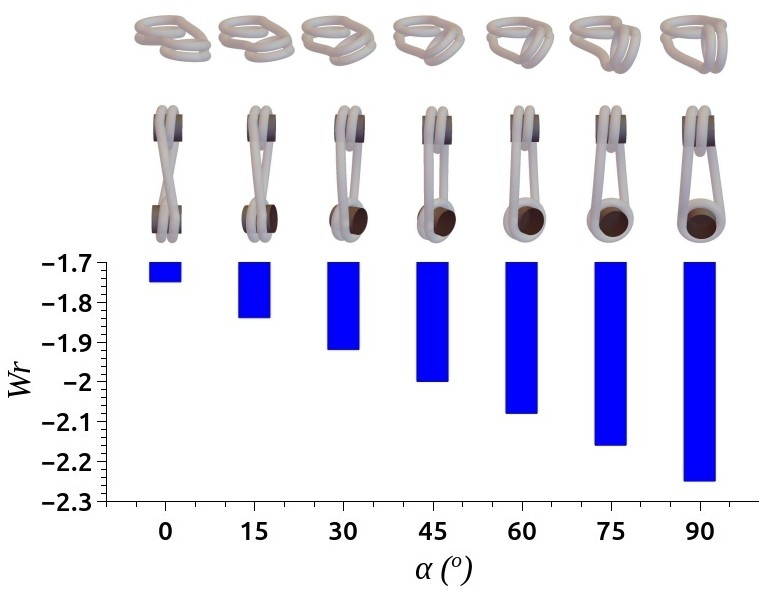}\\
\end{figure}

Figure 3 shows the variation in energy and DNA configuration of a torsionally relaxed, $360$-bp DNA minicircle with two nucleosomes, each 
wrapping $1.75$ turns of DNA. We again let $\alpha$ change from $0^\circ-90^\circ$ at $15^\circ$
increments. The molecular images are shown from the same view points\textemdash i.e., top, side, and front\textemdash used in Figs. 1 and 2.
 As expected, the magnitudes of the writhing numbers are greater than those found in Fig. 2 for nucleosomes wrapping less DNA. The self 
 crossings that
 give rise to the increased magnitude of $Wr$ are evident from all three molecular views in Fig. 3. The changes in the writhing number 
 with increase of $\alpha$, however, are comparable in magnitude to those found for nucleosomes wrapping less DNA and show a similar 
 monotonic increase in magnitude with $\alpha$. The bending energy also monotonically 
 increases when $\alpha$ increases and all three sets of molecular images clearly show that the smooth connectors become more bent 
 with increase in $\alpha$. 
 
As above, the bending energy per base pair stored in the nucleosomal DNA is constant and the change in total bending energy arises from 
the two connectors (see Table S-I). The change in average bending energies of the two connectors are equivalent for values
$\alpha = 0^\circ - 30^\circ$ due to the symmetry of the system. For $\alpha = 45^\circ - 90^\circ$ the two connectors self intersect and 
their contact forces 
break the symmetry and change the average bending energies.

 The electrostatic energy exhibits a local minimum when $\alpha = 75^\circ$. Here again since the 
  variation of the bending energy is larger than that of the electrostatic energy for the chosen parameters, the total energy follows 
  the bending energy and monotonically increases with $\alpha$. It is again possible that for different values of $\epsilon_r$, $\kappa$, 
  and/or charge neutralization the electrostatic energy will dominate the bending energy and the configuration will adopt a local minimum 
  at $\alpha = 60^\circ$.
  
  \begin{figure}[H]
   \centering
   \caption{\small{Molecular images illustrating the changes in overall folding and graphs of the associated bending $\Psi$ 
   and electrostatic $\Phi$ energies, and the writhing number $Wr$ of torsionally relaxed, $360$-bp DNA minicircles with two evenly spaced 
   nucleosomes, each wrapping 
   $1.75$ turns of DNA, as a function of the angle $\alpha$. See Table S-III for coefficients of the smooth connectors and Table S-IV 
   for numerical values of plotted data.}}
   \includegraphics[width=0.48\textwidth]{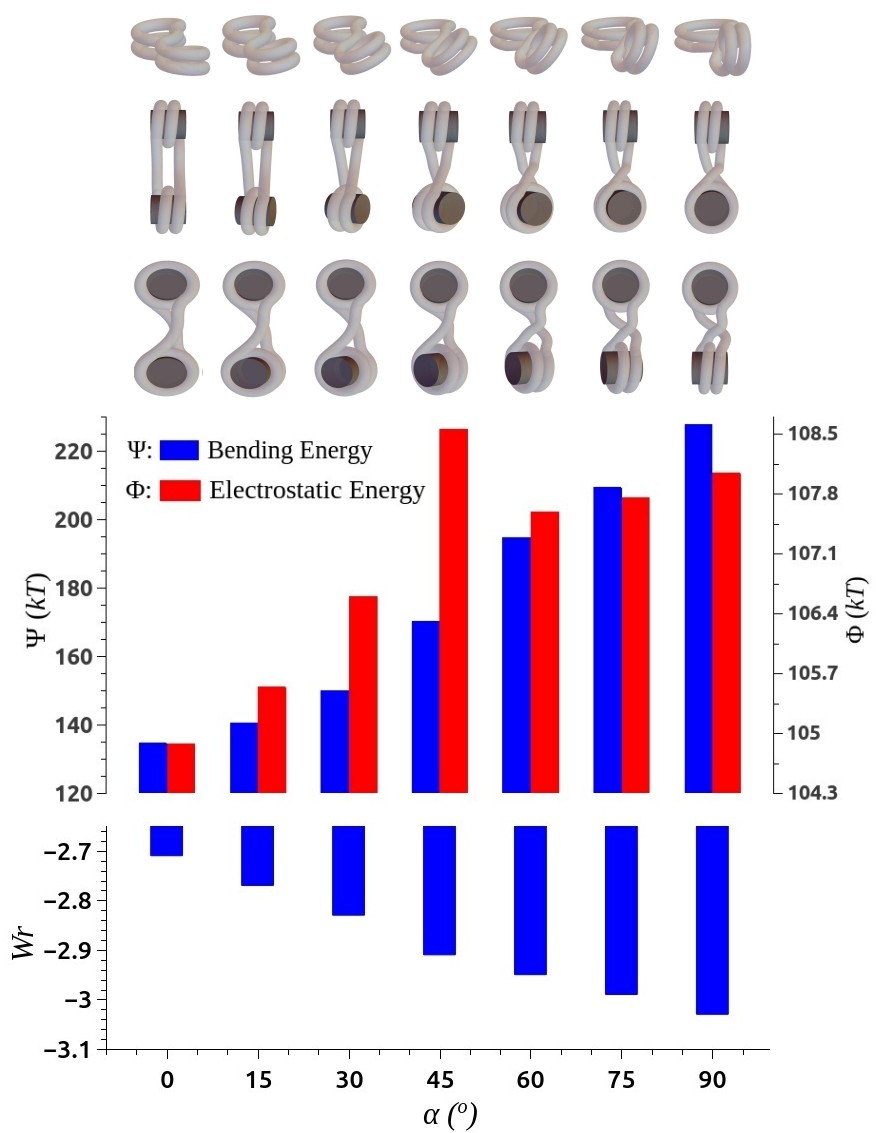}\\
\end{figure}

\section{Discussion}

The minimum-energy configurations of torsionally relaxed DNA minicircles determined in this work show a close resemblance to the shapes of 
chromatin constructs 
reconstituted under low and high salt conditions \cite{Goulet}. Figure 4(a) presents a molecular image of the predicted DNA pathway 
of a $360$-bp minicircle, or so-called dimer, with two evenly spaced nucleosomes, each 
wrapping $1.5$ turns of DNA. The minimum energy configuration of the DNA in the absence of salt, found when the nucleosomes are  oriented 
at an angle $\alpha = 60^\circ$,    
is very similar to the electron microscopic image of the dimer observed under low salt conditions (Fig. 4(b)). 
Figure 4(c) shows a molecular 
image of the predicted DNA pathway of the same dimer with each nucleosome
wrapping $1.75$ turns of DNA. The minimum energy configuration, found at $100mM$ monovalent salt concentration when the 
nucleosomes are oriented 
at an angle $\alpha = 0^\circ$, is very similar to the electron microscopic image of the dimer observed under similar conditions 
(Fig. 4(d)). 
The correspondence of the models with the observed images 
suggests that the addition of salt increases the wrapping of DNA around the nucleosomes from $\sim1.5$ to $\sim1.75$ turns. The model, however, 
does not take account of the 
torsional stress in the reconstituted minicircles, which may contribute to the observed pathways. The experimental constructs are 
negatively supercoiled, with a linking number of $-2$, which could possibly give rise to the observed increased crossings of DNA.
The next step in our studies will be to take the twist of DNA into account and to treat the connector DNA as an elastic rod as opposed 
to a simple curve. This will allow us to determine the twists of successive base pairs and the linking number of the minicircle as a whole. 
The treatment of individual base pairs will also allow us to consider discrete nucleosome structures, such as those available through the 
Protein Data Bank \cite{P.D.Bank}. The boundary value treatment described here can be immediately used to determine the smooth curve that 
connects the 
terminus of one set of coordinates to the start of the next and readily adopted to study the configurational properties of large, multinucleosome 
assemblies such as the simian virus 40 (SV40) minichromosome \cite{Pipas}. The method depends only on the 
values of the coordinates and slopes of the initial and final points. The arc length which connects the two objects to a 
desired length, a scalar, and the slopes, which are vectors, are invariant under 
rotation and translation of the system.

\begin{figure}[H] 
   \centering
   \caption{\small{Predicted configurations of torsionally relaxed $360$-bp DNA minicircles bearing two evenly spaced nucleosomes compared with 
   images of chromatin rings (dimers) of comparable size ($359$-bp) reconstituted under low and high salt conditions. 
  (a) Predicted DNA pathway, in the absence of salt, with nucleosomes wrapping $1.5$ 
  turns of DNA and oriented at an angle $\alpha = 60^\circ$ 
  with respect to one another;
  (b) electron microscopic image of the corresponding dimer at low salt; (c) predicted DNA pathway, with 
  each nucleosome wrapping $1.75$ 
   turns of DNA and oriented at an angle $\alpha = 0^\circ$ under $100mM$ monovalent salt concentration;
   (d) electron microscopic image of the dimer in the presence of $100mM$ NaCl. Images in (a) and (c) viewed along direction angles 
   $\{0.4,-1.5,3.0\}$ and $\{0.2,-0.8,3.3\}$, respectively. Figures (b) and (d) were reproduced with 
   permission from Elsevier (Ref.\cite{Goulet}, Fig. 5(d, e)).}}
   \includegraphics[width=0.45\textwidth]{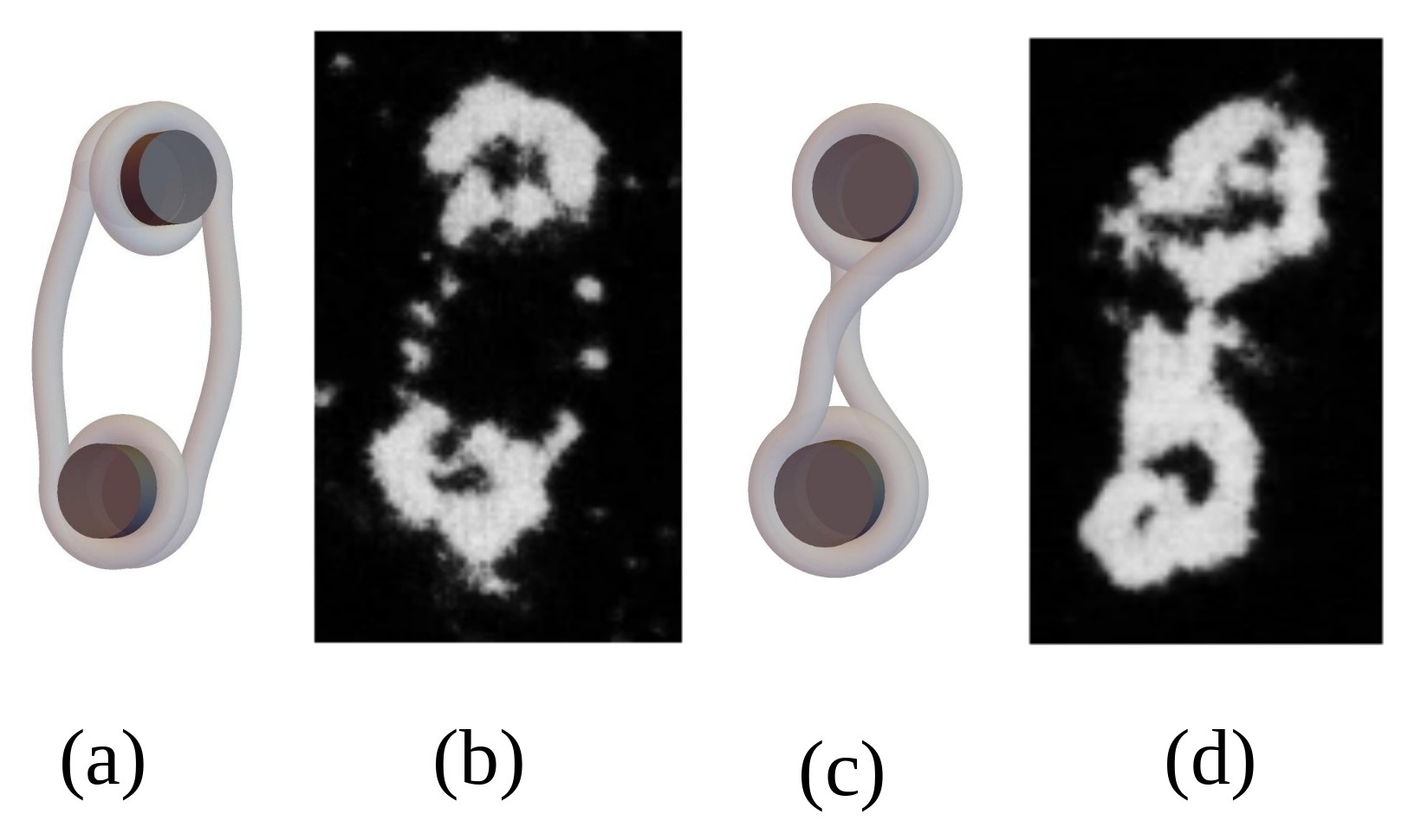}\\
\end{figure}

\section*{ACKNOWLEDGEMENT}
This work was partially supported by the U.S. Public Health Service under research grant GM34809.

\newpage

\appendix*
\section*{Supporting Information}
\begin{center}
 \textbf{\large New smoothly connecting open curves for modeling nucleosome-decorated DNA } 
 \\
  Seyed A. Sabok-Sayr$^1$ and Wilma K. Olson$^{2,\; 3}$\\
     \textit{$^1$Department of Physics and Astronomy, \\ Rutgers University, Piscataway, NJ, U.S.A.} \\
     \textit{$^2$Department of Chemistry and Chemical Biology,}\\
    
\textit{$^3$Center for Quantitative Biology,\\ Rutgers University, Piscataway, NJ, U.S.A.}

\end{center}

\newpage

\section*{Overview}
The supporting information includes four tables respectively listing: the values of the average bending energies per base pair for 
connectors and nucleosomes comprising a torsionally relaxed, $360$-bp DNA minicircle with two evenly spaced nucleosomes, 
each wrapping either $1.5$ or $1.75$ turns of DNA (Table S-I); the values of $d$, $\Psi$, $\Phi$, and $Wr$ plotted in Figs. 1 and 2 
(Table S-II); the values of the coefficients $(a_{\mu k}, \mu=0-4, k=1-2)$ describing the two (or four) smooth connectors for 
minicircles containing two nucleosomes, each with $1.75$ turns of DNA, and oriented at different values of $\alpha$ (Table III); and
the values of $d$, $\Psi$, $\Phi$, and $Wr$ plotted in Fig. 3 (Table S-IV).

Note that the boundary conditions in Table III
for $\alpha = 0^\circ - 30^\circ$ do not introduce self crossings and lead to unique values for $(a_{\mu k}, \mu=0-4, k=1-2)$.
For  $\alpha = 45^\circ - 90^\circ$ the 
smooth connectors begin to self intersect and, as explained in the text, a set of new boundary conditions for the coordinates and slope at the 
contact point has been implemented. This results in two parts for each connector and therefore we have two sets of 
coefficients for each connector.
The reported values of $d$ are the minimum distances between nucleosome centers that satisfy the boundary conditions.  

\newpage

\begin{table}[H]
    \centering
      \caption*{\small{TABLE S-I: Values of the average bending energy per base pair for connectors and 
      nucleosomes comprising torsionally relaxed, $360$-bp DNA minicircles with two evenly spaced nucleosomes, each 
      wrapping either $1.5$ or $1.75$ turns of DNA, as a function of $\alpha$.$^\dagger$}}
      \begin{tabular}{|c|c|c|c|c|}
      
       \hline
       \multicolumn{5}{|c|}{1.5 Turns} \\

       \hline
       \textbf{$\alpha\;(^\circ$)} & \textbf{$E_1 (kT)$} & $E_2(kT)$ & $E_f(kT)$  & $E_s(kT)$ \\
   
       \hline
       $0$  & $0.17$   & $0.17$   & $0.44$   & $0.44$\\
       $15$ & $0.10$   & $0.10$   & $0.44$   & $0.44$\\
       $30$ & $0.06$   & $0.06$   & $0.44$   & $0.44$\\
       $45$ & $0.04$   & $0.04$   & $0.44$   & $0.44$\\
       $60$ & $0.03$   & $0.03$   & $0.44$   & $0.44$\\
       $75$ & $0.04$   & $0.04$   & $0.44$   & $0.44$\\
       $90$ & $0.06$   & $0.06$   & $0.44$   & $0.44$\\
       \hline
       \multicolumn{5}{|c|}{1.75 Turns} \\
       \hline
       $0$  & $0.16$   & $0.16$   & $0.44$   & $0.44$\\
       $15$ & $0.23$   & $0.23$   & $0.44$   & $0.44$\\
       $30$ & $0.33$   & $0.36$   & $0.44$   & $0.44$\\
       $45$ & $0.54$   & $0.63$   & $0.44$   & $0.44$\\
       $60$ & $0.87$   & $0.88$   & $0.44$   & $0.44$\\
       $75$ & $0.92$   & $1.18$   & $0.44$   & $0.44$\\
       $90$ & $0.97$   & $1.56$   & $0.44$   & $0.44$\\
       \hline
      \end{tabular}
      \captionsetup{justification=justified, singlelinecheck=off}
      \caption*{\footnotesize{$^\dagger$ $E_1$ and $E_2$ refer respectively to the average bending energies for connectors 1 and 2, and 
      $E_f$, $E_s$ to those of the first and second nucleosomes.}}

\end{table}

\newpage

\begin{table}[H]
   \begin{center}

      \caption*{\small{TABLE S-II: Calculated values of the minimum separation distances $d$, bending energies $\Psi(kT)$, electrostatic energies $\Phi$, and writhing 
      numbers $Wr$ of a $360$-bp DNA minicircle with two evenly spaced nucleosomes, each wrapping $1.5$ turns of DNA, as a function of $\alpha$.}}

      \label{tab:table2}
      \begin{tabular}{|c|c|c|c|c|c|}

       \hline
         \multirow{2}{*}{\textbf{$\alpha\;(^\circ$)}} & \multirow{2}{*}{\textbf{$d$($\si{\AA}$)}}  &  \multirow{2}{*}{\textbf{$\Psi(kT)$}}  
         &  $10 mM$  & $100 mM $   & \multirow{2}{*}{\textbf{$Wr$}}    \\
        &  &   & \textbf{$\Phi(kT)$}  & \textbf{$\Phi(kT)$}   &  \\
   
       \hline
       $0$  & $200.94$   & $124.32$   & $259.47$   & $119.74$   & $-1.75$ \\
       $15$ & $202.64$   & $116.57$   & $257.35$   & $119.28$   & $-1.84$  \\
       $30$ & $203.66$   & $110.97$   & $255.82$   & $118.96$   & $-1.92$  \\
       $45$ & $203.66$   & $108.26$   & $254.72$   & $118.73$   & $-2.00$  \\
       $60$ & $203.66$   & $107.49$   & $254.59$   & $118.63$   & $-2.08$  \\
       $75$ & $202.64$   & $108.80$   & $254.90$   & $118.60$   & $-2.16$  \\
       $90$ & $200.94$   & $110.98$   & $255.82$   & $118.64$   & $-2.25$  \\
       \hline
      \end{tabular}

   \end{center}

\end{table}

\newpage

\begin{table}[H]
\hfil           
   \begin{turn}{90}
   \begin{minipage}{1.7\linewidth}
   \begin{center}
      \caption*{\small{TABLE S-III: Minimum separation distances and coefficients$^\dagger$ of smooth connectors between two evenly spaced nucleosomes, each 
      with $1.75$ turns of DNA, on a $360$-bp minicircle as a function of the orientation angle $\alpha$.$^\dagger$}}
      \scalebox{0.72}{%
      \label{tab:table1}
      \begin{tabular}{|c|c|c|c|c|c|c|c|c|c|c|c|}
       \hline
       \textbf{$\alpha\;(^\circ$)} & \textbf{$d$($\si{\AA}$)} & \textbf{$a_{41}(\si{\AA^{-3}})$} & \textbf{$a_{31}(\si{\AA^{-2}})$} & \textbf{$a_{21}(\si{\AA^{-1}})$} & \textbf{$a_{11}$} &
       \textbf{$a_{01}(\si{\AA})$} & \textbf{$a_{42}(\si{\AA^{-3}})$} & \textbf{$a_{32}(\si{\AA^{-2}})$} & \textbf{$a_{22}(\si{\AA^{-1}})$} & \textbf{$a_{12}$} &
       \textbf{$a_{02}(\si{\AA})$}   \\  
       \hline
        $0$   & $180.88$ & $1.14\times10^{-6}$ & $4.82 \times 10^{-4}$  &  $6.64 \times 10^{-2}$ & $3.84$  &  $98.83$ &
               $1.14\times10^{-6}$ & $3.45 \times 10^{-4}$  &  $2.94 \times 10^{-2}$ & $-3.95 \times 10^{-2}$  &  $-50.47$\\
        $15$ & $178.84$ & $1.24\times10^{-6}$ & $5.04 \times 10^{-4}$  &  $6.71 \times 10^{-2}$ & $3.80$  &  $101.48$ & 
               $1.31\times10^{-6}$ & $4.07 \times 10^{-4}$  &  $3.74 \times 10^{-2}$ &  $3.32 \times 10^{-1} $ &  $-49.08$ \\
        $30$ & $177.14$ &  $1.25\times10^{-6}$& $4.94 \times 10^{-4}$  &  $6.36 \times 10^{-2}$ & $3.51$  &  $97.40$ & 
                $1.54\times10^{-6}$& $4.93 \times 10^{-4}$  &  $4.88 \times 10^{-2}$ & $9.17 \times 10^{-1}$& $-41.35$   \\
                \hline
        \multirow{2}{*}{$45$} & \multirow{2}{*}{$171.36$} & $-5.52\times10^{-6}$ & $-1.23\times10^{-3}$  &  $-8.82\times10^{-2}$  &  $-1.82$ &  $32.02$ &
               $4.81\times10^{-6}$ & $1.91\times10^{-3}$  &  $2.77\times 10^{-1}$ & $16.93$  & $370.44$  \\
               & & $1.05\times10^{-5}$ & $4.51\times10^{-3}$ & $6.98 \times 10^{-1}$ & $46.80$ & $1173.99$ & $-5.63\times10^{-7}$ & $2.57\times10^{-5}$ & $1.55\times10^{-2}$ & $1.36\times 10^{-1}$ & $-44.89$\\
                \hline
        \multirow{2}{*}{$60$} & \multirow{2}{*}{$168.30$} &  $1.83\times10^{-5}$ & $4.43\times10^{-3}$  &  $3.74\times10^{-1}$ & $13.13$  &  $194.95$ &  
               $-6.03\times10^{-6}$ & $-3.27\times10^{-3}$  &  $-6.39\times10^{-1}$ & $-53.62$  & $-1623.27$   \\
               & & $1.58\times10^{-5}$ & $ 6.94\times10^{-3}$ & $1.12$ & $78.49$ & $2068.68$ & $1.51\times10^{-5}$ & $3.18\times10^{-3}$ & $2.43\times10^{-1}$ & $7.17$ & $36.83$ \\
                \hline
        \multirow{2}{*}{$75$} & \multirow{2}{*}{$166.60$} & $-3.03\times10^{-6}$  & $1.04\times10^{-4}$  &  $5.86\times10^{-2}$ & $3.32$  &  $79.69$ &  
               $-2.71\times10^{-6}$ & $-1.63\times10^{-3}$  &  $-3.33 \times 10^{-1}$ & $-28.52$  & $-858.75$  \\  
               & & $-9.81\times10^{-6}$ & $-3.92\times10^{-3}$ & $-5.62\times10^{-1}$ & $-33.24$ & $-653.36$ & $1.95\times10^{-5}$ & $4.10\times10^{-3}$ & $3.15 \times10^{-1}$ & $9.58$ & $71.79$ \\
                \hline
               \multirow{2}{*}{$90$} & \multirow{2}{*}{$162.18$} & $7.17\times10^{-6}$ & $2.35\times10^{-3}$  &  $2.31\times10^{-1}$ & $8.58$  &  $127.18$ &         
                             $-7.40\times10^{-6}$& $-3.75\times10^{-3}$  & $-6.86\times10^{-1}$    & $-54.04$    & $-1534.33$ \\
               & &$-1.23\times10^{-5}$ & $-4.85\times10^{-3}$&$-6.93\times10^{-1}$ &$-41.33$ &$-837.74$ & $2.91\times10^{-5}$ &  $5.84\times10^{-3}$  & $4.25\times10^{-1}$     & $12.67$      &$112.20$\\ 
       \hline
      \end{tabular}}
      \captionsetup{justification=justified, singlelinecheck=off}
      \caption*{\footnotesize{$^\dagger$The first subscript of each coefficient is its order (Eq.$(1)$) and the second one refers to the order of connectors.
      In the second subscript, values $1$ and $2$ refer respectively to the connectors joining the terminus of the second nucleosome 
      to the start of the first nucleosome and the terminus of the first nucleosome to the start of the second.}}
   \end{center}
  \end{minipage}
  \end{turn}  
\end{table}

\newpage

\begin{table}[H]

    \centering
      \caption*{\small{TABLE S-IV: Calculated values of the minimum separation distances $d$, bending energies $\Psi$, electrostatic energies $\Phi$, and 
      writhing numbers $Wr$ of a $360$-bp DNA minicircle with two 
      evenly spaced nucleosomes, each with wrapping $1.75$ turns of DNA, as a function of $\alpha$.}}

      \begin{tabular}{|c|c|c|c|c|c|}

       \hline
         \multirow{2}{*}{\textbf{$\alpha\;(^\circ$)}} & \multirow{2}{*}{\textbf{$d$($\si{\AA}$)}}  &  \multirow{2}{*}{\textbf{$\Psi(kT)$}}  
         &  $10 mM$  & $100 mM $   & \multirow{2}{*}{\textbf{$Wr$}}    \\
        &  &   & \textbf{$\Phi(kT)$}  & \textbf{$\Phi(kT)$}   &  \\
   
       \hline
       $0$  & $180.88$   & $134.65$ & $234.28$  & $104.88$  & $-2.71$\\
       $15$ & $178.84$   & $140.43$ & $247.72$  & $105.54$  & $-2.77$\\
       $30$ & $177.14$   & $150.11$ & $252.98$  & $106.60$  & $-2.83$\\
       $45$ & $171.36$   & $170.39$ & $259.07$  & $108.55$  & $-2.91$\\
       $60$ & $168.30$   & $194.67$ & $257.23$  & $107.59$  & $-2.95$\\
       $75$ & $166.60$   & $209.34$ & $258.55$  & $107.76$  & $-2.99$\\
       $90$ & $162.18$   & $227.78$ & $260.86$  & $108.04$  & $-3.03$\\
       \hline
      \end{tabular}
\end{table}
\end{document}